# n-vicinity method for Ising Model with long-range interaction

**Inna Kaganowa, Boris Kryzhanovsky and Leonid Litinskii**

Center of Optical Neural Technologies, Scientific Research Institute for System Analysis RAS, Moscow, Russia

E-mail: imkaganova@gmail.com, kryzhanov@mail.ru, litin@mail.ru

**Abstract**

The previously developed n-vicinity method allows us to calculate accurately critical values of inverse temperatures for Ising models with short-range interaction. We generalize the method to the case of long-range interactions in spin systems and obtain theoretical formulas for the inverse temperatures in terms of the spin interaction constants. The comparison of our theoretical estimates with computer simulations for the two- and three-dimensional Ising models shows that the larger the dimension of the problem the better their agreement.

## 1. Introduction and problem formulation

Methods of statistical physics allows the scientists to solve problems in various fields of science such as material science and econophysics, the neural network theory and optimization problems, and so on [1]. The free energy is the core of all statistical physics problems. It helps to calculate the most important macroscopic characteristics of various systems. To calculate the free energy of a spin system it is necessary to know the energy distribution of all $2^N$ states of the system, where $N$ is the number of spins. As a rule, the energy state distribution is unknown.

Previously we developed the n-vicinity method, which is a universal method for calculation of the free energy of a large spin system ($N \gg 1$). To apply the method we have to fix a spin configuration $\mathbf{s}_0$ as initial and to distribute all other spin configurations between its $n$-vicinities $\Omega_n$, where $\Omega_n$ consists of the configurations that differ from the initial configuration $\mathbf{s}_0$ in opposite signs of n binary coordinates. Here $n$ varies from 0 to $N$ and the number of configurations in the $n$-vicinity $\Omega_n$ is equal to $C_N^n$.

Even if we do not know the energy distribution of the states belonging to $\Omega_n$, we can calculate exactly its mean value $E_n$ and the variance $\sigma_n^2$ in terms of the parameters of the connection matrix $\mathbf{T} = (T_{ij})_{i,j=1}^{N}$ of the spin system and the initial configuration $\mathbf{s}_0$ [2], [3]. Our computer simulations show (and we can justify this statement using the central limit theorem) that when $N$ is large, the Gaussian distribution with the mean $E_n$ and the variance $\sigma_n^2$ is a satisfactory approximation for the energy state distribution. The only approximation of the n-vicinity method is the replacement of the real distribution from $\Omega_n$ by the Gaussian distribution with the "correct" mean value $E_n$ and the variance $\sigma_n^2$.

We obtain the energy distribution of all $2^N$ states summarizing "the partial" Gaussian distributions for all the vicinities $\Omega_n$. After tending $N$ to infinity, we pass from the summation over $n$ to integration over a variable $m = 1 - 2n/N$ that characterizes the magnetization of the state. Using a standard procedures, we obtain the equation of state with the aid of which we calculate the free energy for the given value of the inverse temperature $\beta = 1/T$. For spin systems discussed below, the equation of state is very simple:

$$\frac{1}{2m}\ln\frac{1+m}{1-m} = \gamma b\left[1 - b + bm^2\right], \tag{1}$$

where $m \in [-1,1]$ is a magnetization and other parameters entering this equation are

$$b = \beta \frac{\sum_{ij} T_{ij}^2}{\sum_{ij} T_{ij}} \ , \ \gamma = \frac{\left(\sum_{ij} T_{ij}\right)^2}{N \sum_{ij} T_{ij}^2} . \tag{2}$$

Here $b$ is a normalized inverse temperature $\beta$ and the dimensionless parameter $\gamma$ characterizing interaction of each spin with its surrounding is in fact the effective number of neighbors of the given spin.

To test the equations (1) and (2) we used a standard Ising model on hypercube [4]. In this case, the spins are at the nodes of a hypercube lattice and each spin interacts with its nearest neighbors only; all the nonzero interaction constants are the same. In paper [3], we showed that for the standard model, with the aid of our approach it was possible to derive a simple analytical expression for the critical value of the inverse temperature in terms of the lattice dimension. We found that the obtained expression describes accurately the results of computer simulations for this Ising model when the lattice dimension was in the range from 3 to 7. We need these results in what follows and recall them in Section 2.

In the present publication, we go beyond the framework of the standard Ising model, and take into account the interactions with not the nearest neighbors only but with next neighbors too. We suppose that the interaction constant with the nearest neighbors is equal to 1, and interaction constants with more distant neighbors vary from 0 to 1. In Section 3, we discuss a planar Ising model. In this case, we have a plane square lattice where spins are at the vertices of the squares and account for interactions with the next neighbors. In Section 4, we examine a cubic model where $d = 3$ and take into account interactions not only with the next neighbors but also with next-next neighbors. In these problems, we can use the $n$-vicinity method, which allows us to express the inverse temperature in terms of the interaction parameters. In Section 5 are discussion and conclusions.

**2. Interaction with nearest neighbors**

For what follows it worth mentioning that if the hypercube dimension is $d$, the number of the nearest neighbors of each spin is equal to $q = 2d$. It is known that when the hypercube dimension $d > 1$ and we take into account interactions with the nearest neighbors only, the Ising models demonstrate phase transitions of the second kind but there are no phase transitions of the first kind [4]. In [3] we showed that for the equation of state (1) this was true when the parameter

$$\gamma > 16/3. \tag{3}$$

For such values of $\gamma$ we can write the explicit expression for the critical value of the inverse temperature:

$$\beta_c = \frac{\sum_{i,j} T_{ij}}{2 \cdot \sum_{i,j} T_{ij}^2} \left(1 - \sqrt{1 - 4/\gamma}\right), \tag{4}$$

with $\gamma$ defined by Eq. (2). When we account for the interactions with the nearest neighbors only, Eq. (4) is particularly simple. In this case, the value of the parameter $\gamma$ is equal to the number of the nearest neighbors $q$:

$$\gamma = q = 2d, \tag{5}$$

and the expression (4) is also shorter

$$\beta_c = \frac{1 - \sqrt{1 - 2/d}}{2}. \tag{6}$$

Let us note that with regard of Eq. (3) the equations (5) and (6) are valid only if the hypercube dimension $d > 8/3$ and this makes it impossible to use them when analyzing the planar Ising model ($d = 2$). However, for the dimensions $d = 3, 4, 5, 6, 7$ this inequality fulfills and for all such lattices the critical inverse temperature defined by Eq. (6) is in good agreement with the results of computer simulations, see [3, 5, 6, and 7] and Table 1.

Table 1. Comparison of estimate (6) and computer simulations

| $d$ | 3 | 4 | 5 | 6 | 7 |
|---|---|---|---|---|---|
| $\beta_c$ (6) | 0,2113 | 0,1464 | 0,1127 | 0,0918 | 0,0774 |
| Experiment | 0,2216 | 0,1489 | 0,1139 | 0,0923 | 0,0777 |

Let us note, the larger the hypercube dimension $d$ the better agreement of the theory and the results of computer simulations. When $d = 3$ the relative difference between the theory and computer simulations is about 2.5%, but when $d$ increases from 3 to 7 the relative difference decreases up to 0,25%. We can say that the predictive ability of our simple analytic expression (6) is rather high.

### 3. Planar Ising model with account for next neighbors

Now we will show that a ferromagnetic interaction with next neighbors allows us to increase the parameter $\gamma$ in such a way that it would satisfy the inequality (3). This means the ability to use the *n*-vicinity method when analyzing the planar Ising model. In part, we published the results presented below in [8].

In the case of the planar square lattice, each spin has four nearest neighbors at the distance equal to the side of the square; it also has four next neighbors at the distance equal to the diagonal of the elementary square. We suppose that the interaction constants with the nearest neighbors $J_1 = J_2 = 1$ and $J_3$ is the interaction constant with the next neighbors. Then from Eq. (2) we obtain

$$\gamma = 4\frac{(1+J_3)^2}{1+J_3^2}. \tag{7}$$

If $J_3 > 3 - 2\sqrt{2} \approx 0.2$, the value of $\gamma$ calculated according Eq. (7) satisfies Eq. (3). Consequently, the *n*-vicinity method is applicable and the critical inverse temperature is

$$\beta_c = \frac{1+J_3-\sqrt{2J_3}}{2(1+J_3^2)}. \tag{8}$$

We verified the obtained formulas using computer simulations. For different values of the parameter $J_3$ we generated the $N = 10^4$ dimensional connection matrix that accounted for the next-neighbors and then used the Monte-Carlo method to calculate the mean energy and its variance $\sigma_E^2(\beta)$ for each value of the inverse temperature $\beta = 1/T$. Since the second derivative of the free energy was proportional to the variance $\sigma_E^2(\beta)$, we considered the points of singularities on the graphs of the functions $\sigma_E^2(\beta)$ as the critical values of the inverse temperature. We present the results of our experiments in Table 2, where the first column shows the values of the parameter $J_3$, the corresponding values of $\gamma$ are in the second column, the experimental values of the critical temperature $\beta_c^{(ex)}$ are in the third column, and it's theoretical estimates $\beta_c$ (8) are in the fourth column. In the fifth column, we show the values of $\beta_c$ multiplied by a normalization factor 1.3.

Comparing the third, fourth, and fifth columns, we see that while the experimental values $\beta_c^{(ex)}$ and their theoretical estimates $\beta_c$ differ noticeably, the values $\beta_c^{(ex)}$ and $1.3\beta_c$ are almost equal for all the examined values of the parameter $J_3$. We characterize the difference between them by the ratio

$$\delta = \frac{1.3\beta_c - \beta_c^{(ex)}}{1.3\beta_c + \beta_c^{(ex)}} \cdot 100\%,$$

(the last column of Table 2). It is clear that the difference between these two estimates is within one percent.

Table 2. Experimental ($\beta_c^{(ex)}$) and theoretical ($\beta_c$) values
of critical inverse temperatures for different $J_3$.

| $J_3$ | $\gamma$ | $\beta_c^{(ex)}$ | $\beta_c$ | $1.3\beta_c$ | $\delta$,% |
|---|---|---|---|---|---|
| 0,200 | 5,538 | 0,3460 | 0,2729 | 0,3547 | 1,24 |
| 0,268 | 6,000 | 0,3220 | 0,2500 | 0,3250 | 0,46 |
| 0,369 | 6,600 | 0,2900 | 0,2243 | 0,2916 | 0,28 |
| 0,500 | 7,200 | 0,2620 | 0,2000 | 0,2600 | -0,38 |
| 0,630 | 7,608 | 0,2355 | 0,1817 | 0,2361 | 0,13 |
| 0,750 | 7,840 | 0,2200 | 0,1681 | 0,2185 | -0,34 |
| 0,850 | 7,948 | 0,2066 | 0,1585 | 0,2061 | -0,12 |
| 1,000 | 8,000 | 0,1900 | 0,1464 | 0,1904 | 0,11 |

The figure 1 shows that the experimental points $\beta_c^{(ex)}$ are almost located on the graph of the curve $1.3\beta_c$. An isolated point at the ordinate axis corresponds to the planar Ising model with the interaction between the nearest neighbors only. In the end of 40-th Lars Onsager showed that in this case the exact value of the inverse temperature was $\beta_c^{(0)} = 0.4407$. We cannot continue our graph to this point since when $J_3 < 0.2$ then $\gamma < 16/3$. However, we see that our graph directed to this point.

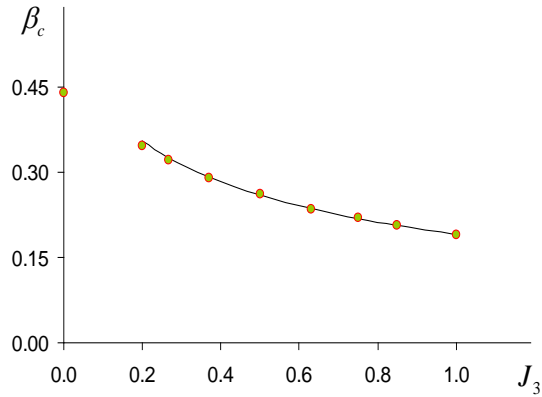

Fig. 1. Critical values of the inverse temperature:
circles are experimental values and solid line is $1.3\beta_c$.

In the last Section, we discuss the meaning of the normalization factor 1.3. In any case, the ability to match the experimental values and the theoretical curve over the large interval of the values of $J_3$ is noteworthy.

**4. Cubic Ising model with account for second and third neighbors**

When analyzing the cubic lattice, we take into account interactions not only with the next neighbors, but also with the next-next neighbors. Then each spin has 6 nearest neighbors, 12 next neighbors, and 8 next-next neighbors. As above, we suppose that all the interactions between the nearest neighbors are equal, $J_1 = J_2 = J_3 = 1$. We successively change the interactions with the next neighbors $J_4$ and with the next-next neighbors in the interval [0,1]. Then the formulas (2) and (4) takes the form

$$\gamma = \frac{2\cdot(3+6J_4+4J_5)^2}{(3+6J_4^2+4J_5^2)} \,, \quad \beta_c = \frac{3+6J_4+4J_5}{2\cdot(3+6J_4^2+4J_5^2)}\left(1-\sqrt{1-\frac{4}{\gamma}}\right). \tag{9}$$

We performed the computer simulations in the same way as described above. Namely, for fixed values of $J_4$ and $J_5$ we generated connection matrices, then we changed the inverse temperature $\beta \sim 1/T$ and for each its value calculated the mean energy and the variance $\sigma_E^2(\beta)$. The point of singularity on the graph of the function $\sigma_E^2(\beta)$ defines the critical value of the inverse temperature for given $J_4$ and $J_5$.

In Fig. 2, we show the graphs of the normalized variance $\sigma^2(\beta)/\sigma_0^2$ for different values of $J_4$ and $J_5 \equiv 0$ when the dimension of the connection matrices is $N = 8 \cdot 10^3$. The normalization factor $\sigma_0^2 = 3 \cdot (1 + 2 \cdot J_4^2)$ is equal to a reduced sum of the squares of matrix elements. It allows us to present all the graphs in one figure, in other way, it would be impossible to do this.

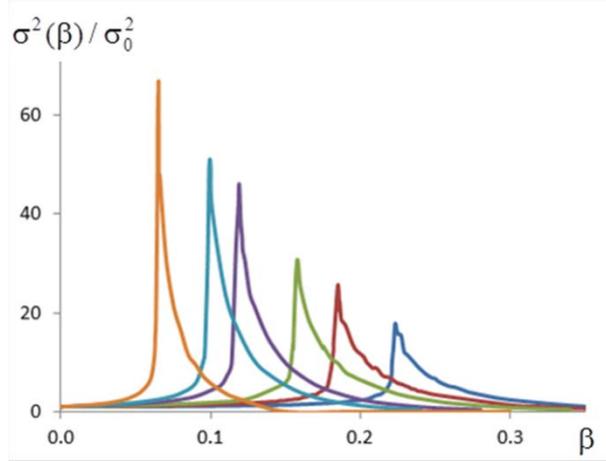

Fig. 2. Graphs of normalized variance for
$J_4 = 0$, 0.0815, 0.16227, 0.35425, 0.5, and 1 (from right to left).

Let us clarify that we searched for the maximum points $\beta_{max}$ on the graphs of the functions $\sigma_E^2(\beta)$, but not the points of discontinuity. When the dimension of the matrix is finite, the graphs have no discontinuities. We interpreted the obtained $\beta_{max}$ as the critical values of the inverse temperature $\beta_c^{(ex)}$ supposing that $\beta_c^{(ex)} \approx \beta_{max}$. For $N = 8 \cdot 10^3$, we found that the coefficient relating the experimental estimates $\beta_{max}$ and the theoretical values $\beta_c$ defined by Eq. (9) is equal to 1.11:
$\beta_{max} \approx 1.11 \cdot \beta_c$

We performed our final simulations for the matrices of the dimension $N = 24 \cdot 10^4$. At first, we changed the parameter $J_4$ and then $J_5$ too. In this case, we found that the consistency between the theoretical and experimental estimates was better: $\beta_{max} \approx 1.08 \beta_c$. We show the results of this experiment in Table 3. The values of the interaction parameters $J_4$ and $J_5$ are in the first and second columns, respectively, and the corresponding values of $\gamma$ calculated according Eq. (9) are in the third column. It is noteworthy to mention that at first, we set $J_5 = 0$ and change $J_4$ from zero to one and then, leaving $J_4 = 1$, we change $J_5$ in the same range. We recall that $\gamma$ defines the effective number of neighbors of the given spin. We see that this number increases monotonically from 6 to 26. The values of $\beta_c$ obtained with the aid of Eq. (9) are in the fourth column, the values of $1.08\beta_c$ are in the fifth column, and the experimental estimates of the critical inverse temperature $\beta_{max}$ are in the sixth column. In the last column are the relative differences between $1.08\beta_c$ and $\beta_{max}$. We see that the relative errors are about fractions of a percent.

Table 3. Comparison of experimental ($\beta_{max}$) and theoretical ($\beta_c$) values of inverse temperature for different $J_4$ and $J_5$.

| $J_4$ | $J_5$ | $\gamma$ | $\beta_c$ (9) | $1.08 \cdot \beta_c$ | $\beta_{max}$ | $\delta$, % |
|---|---|---|---|---|---|---|
| 0 | 0 | 6,000 | 0,21132 | 0,22823 | 0,22201 | -1,38 |
| 0,08114 | 0 | 8,000 | 0,16800 | 0,18144 | 0,18080 | -0,18 |
| 0,16228 | 0 | 10,000 | 0,14181 | 0,15315 | 0,15330 | 0,05 |
| 0,25000 | 0 | 12,000 | 0,12234 | 0,13212 | 0,13343 | 0,49 |
| 0,35425 | 0 | 14,000 | 0,10574 | 0,11420 | 0,11640 | 0,96 |
| 0,50000 | 0 | 16,000 | 0,08932 | 0,09646 | 0,09800 | 0,79 |
| 0,70000 | 0 | 17,455 | 0,07396 | 0,07987 | 0,08031 | 0,27 |
| 0,85000 | 0 | 17,890 | 0,06563 | 0,07088 | 0,07070 | -0,12 |
| 1 | 0 | 18,000 | 0,05904 | 0,06376 | 0,06400 | 0,18 |
| 1 | 0,06375 | 19,000 | 0,05721 | 0,06179 | 0,06190 | 0,09 |
| 1 | 0,13070 | 20,000 | 0,05543 | 0,05987 | 0,06014 | 0,23 |
| 1 | 0,20228 | 21,000 | 0,05366 | 0,05796 | 0,05799 | 0,03 |
| 1 | 0,28063 | 22,000 | 0,05187 | 0,05602 | 0,05609 | 0,06 |
| 1 | 0,36935 | 23,000 | 0,05000 | 0,05400 | 0,05398 | -0,02 |
| 1 | 0,47549 | 24,000 | 0,04795 | 0,05179 | 0,05171 | -0,07 |
| 1 | 0,61765 | 25,000 | 0,04549 | 0,04913 | 0,04889 | -0,24 |
| 1 | 1 | 26,000 | 0,04007 | 0,04327 | 0,04310 | -0,20 |

In Fig. 3, we show the dependence of $1.08\beta_c$ on the value of $\gamma$ (see Eq. (9)). As follows from Eq. (9), the parameter $\gamma$ is a nonlinear function of the interaction constants that we varied in the course of our experiment. Consequently, the value of $1.08\beta_c$ also is a nonlinear function of these parameters. In the figure, there is a bending of the curve when $\gamma$ is close to 18. From Table 3 it follows that $\gamma = 18$ corresponds to $J_4 = 1$ and $J_5 = 0$. After that $\gamma$ increases due to increase of $J_5$ only. We see that the experimental points $\beta_{max}$ fit well the complex theoretical curve.

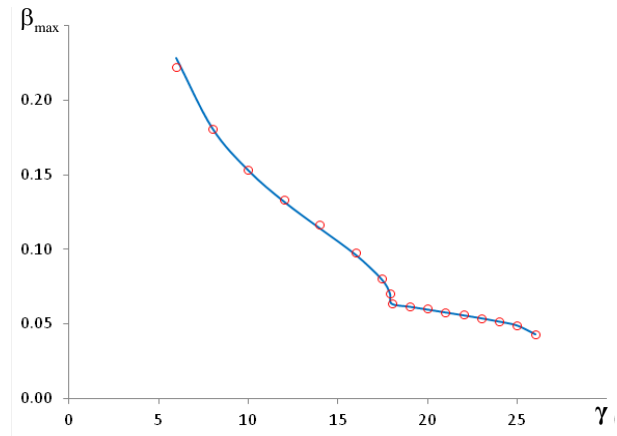

Fig. 3. Critical values of inverse temperature; circles are experimental values $\beta_{max}$, solid line is theoretical curve $1.08\beta_c$.

## 5. Conclusions

We showed that when analyzing the Ising model, the *n*-vicinity method is effective not only if we account for interactions between the nearest neighbors only but also in the cases when the interactions with next and next-next neighbors are significant. The obtained formulas (8) and (9) allow us to express explicitly the critical temperature as function of the interaction constants.

In all the discussed cases, we were able to fit the results of computer simulations with the theoretical curves normalizing them by the factors that were close to one. For the cubic lattice we found that the greater the dimension of the interaction matrix the better the coincidence of the experimental and theoretical estimates. When we increased the dimension of the interaction matrix from $N = 8 \cdot 10^3$ to $N = 24 \cdot 10^4$, the value of the normalizing coefficient decreased from 1.11 to 1.08. We hope that if we increase the number *N* of spins the theoretical estimates (9) would be even closer to the results of computer simulations.

We also hope that the same is also true for the planar Ising model. In this case, for the experiment described above we obtained a rather large normalizing factor 1.3. We think that if we would increase the dimension of the problem by 1-2 orders of magnitude (let us say, from $N = 10^4$ to $N = 10^6$) the value of the normalizing factor would be much closer to one.

**Acknowledgements**
This work was supported by Russian Basic Research Foundation grant # 18-07-00500.